\documentclass[twocolumn, superscriptaddress,showpacs,preprintnumbers,amsmath,amssymb]{revtex4}
\usepackage{graphicx}
\usepackage{bm}

\newcommand{\vect}[1]{\bm{#1}}
\newcommand{\deffig}[4]{
 \begin{figure}
 \includegraphics[scale=#3]{#2}
 \caption{\label{#1} #4}
 \end{figure}
}
\newcommand{\beq}{\begin{equation}}
\newcommand{\eeq}{\end{equation}}
\begin{document}


\title{
Dimer-Quadrupolar Quantum Phase Transition 
in the Quasi-One-Dimensional Heisenberg Model with Biquadratic Interaction
}

\author{Kenji Harada}
\affiliation{
 Department of Applied Analysis and Complex Dynamical Systems, 
 Graduate School of Informatics,
 Kyoto University, Kyoto 606-8501, Japan
}
\author{Naoki Kawashima}
\affiliation{
 Institute for Solid State Physics, University of Tokyo,
 Kashiwa 5-1-5, Chiba 277-8581, Japan
}
\author{Matthias Troyer}
\affiliation{
 Theoretische Physik, Eidgen\"ossische Technische Hochschule Zurich, 
 8093 Zurich, Switzerland
}

\date{\today}


\begin{abstract}
The quasi-one-dimensional $S=1$ Heisenberg antiferromagnet 
with a biquadratic term is investigated at zero temperature
by quantum Monte Carlo simulation.
As the magnitude of the inter-chain coupling is increased, 
the system undergoes a phase transition 
from a spontaneously dimerized phase to a N\'eel ordered 
or spin nematic phase.
The numerical results suggest the possibility of an unconventional 
second-order transition in which 
the symmetry group characterizing one phase is not a 
subgroup of the other.
\end{abstract}

%
%

\maketitle


In the search for novel quantum phases and transitions,
various spin models with interactions beyond the ordinary
bilinear couplings have been investigated.
Among such models, the $S=1$ Heisenberg model with 
biquadratic interaction is known to possess
various interesting non-magnetic ground states, such as
a dimerized and a Haldane phase in one dimension
and a quadrupolar (or spin nematic) phase in two or more dimensions.
In most magnetic systems the biquadratic term
corresponds to a higher order exchange coupling and is
much smaller in magnitude than the bilinear interactions,
bearing no large effects upon the thermodynamic behavior. 
The investigation of this model at large values of the biquadratic term 
was thus mostly of academic interest.
Recently, however, experiments \cite{ColdAtoms} on optically trapped 
ultra-cold atoms opened new possibilities of flexible and controllable 
quantum systems.
In particular, cold atoms in a periodic potential (e.g. an optical lattice),
exhibit a superfluid to Mott-insulating phase transition \cite{OpticalLattice},
which was subsequently studied by quantum Monte Carlo simulation
\cite{BatrouniETAL2002,Kashurnikov2002,Wessel2004,Bergkvist2004}
showing a potential capability of realizing various
strongly-correlated quantum systems.
More closely related to the present study, it was shown 
\cite{Yip,Imambekov} 
that Na atoms trapped in an optically generated periodic potential, 
forming a Mott insulator with one atom per lattice site, 
the hyper-fine $S=1$ spins in the insulating phase 
may be described as an $S=1$ Heisenberg antiferromagnet with
a large biquadratic term, comparable in magnitude to the bilinear term.

\deffig{PhaseDiagram}{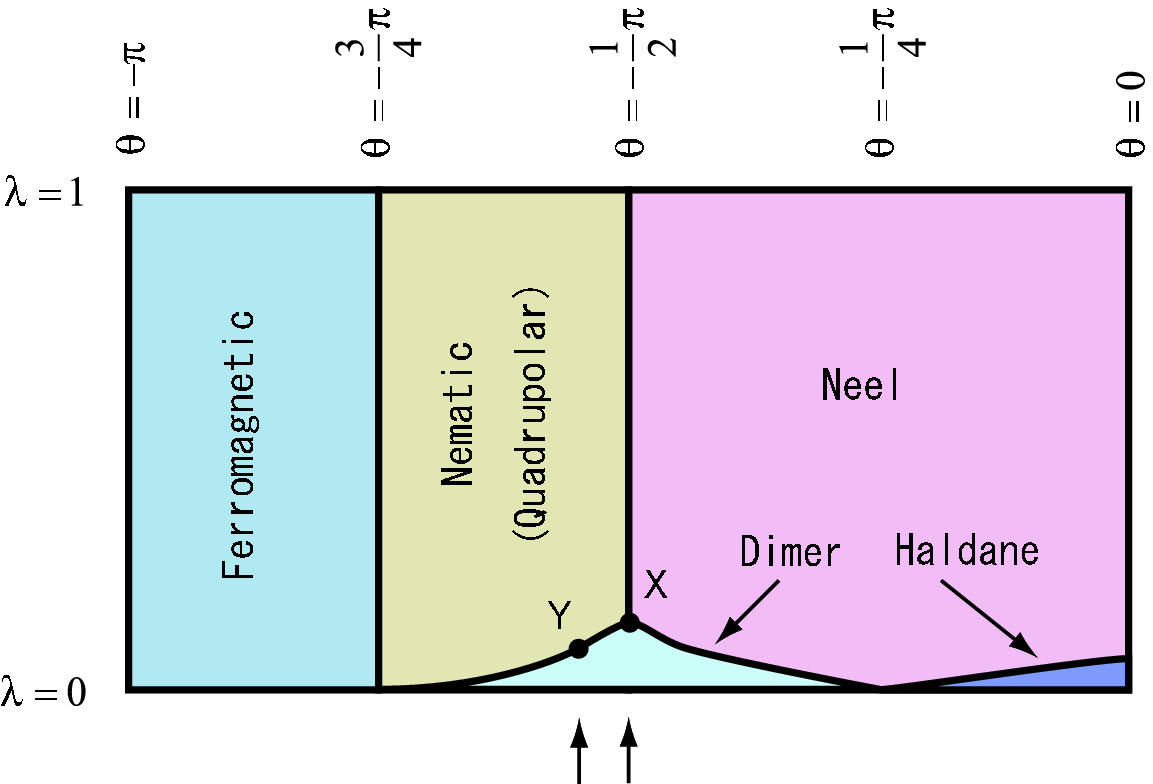}{0.68}
{
 A schematic phase diagram of the $S=1$ Heisenberg antiferromagnet 
 with a biquadratic term.
 Two arrows indicate values of $\theta$ (see text for the definition) 
 at which Monte Carlo simulations are carried out.
}

In this Letter we present the results of 
quantum Monte Carlo simulations of the two-dimensional model 
with strong spatial anisotropy.
When the anisotropy is varied this model exhibits a quantum phase transition 
from a dimerized phase to a spin nematic phase or a N\'eel phase, 
as shown in Fig. \ref{PhaseDiagram}.
The dimerized phase present in one dimension does not survive in
the isotropic two- and three-dimensional model as conjectured by
reference \cite{Yip}, but is stable only in anisotropic quasi-one-dimensional 
systems.
Of particular interest is the quantum phase transition between the dimerized 
and the spin nematic phase, at which we find that
the correlation length is either infinite or at least very large ($\xi \gg 100$).
If the transition is of the second order, 
it must fall into an unconventional class of critical phenomena 
\cite{SenthilVBSF2004,SenthilBSVF2004,SenthilBSVF2005}, 
because the symmetry group characterizing one phase is not 
a sub-group of the other.

The model we consider is
described by the following $S=1$ spin Hamiltonian:
$$
 H = \sum_{(\vect{r},\vect{r}')}
 \left( 
 J^{(L)}_{\vect{r},\vect{r}'} \vect{S}(\vect{r})\cdot\vect{S}(\vect{r}') 
 + 
 J^{(Q)}_{\vect{r},\vect{r}'} (\vect{S}(\vect{r})\cdot\vect{S}(\vect{r}'))^2 
 \right),
$$
where the couplings are parameterized by an angle parameter $\theta$ such that
$J^{(L)}_{ij} = J_{ij} \cos \theta$, and 
$J^{(Q)}_{ij} = J_{ij} \sin \theta$.
We focus on the region $\theta \le 0$ and will specifically 
consider two cases $\theta/\pi = -1/2$ and $-0.55$.
While the model generally has the obvious SU(2) symmetry,
it possesses an enlarged SU(3) symmetry in the case of $\theta = -\pi/2$.

A rough sketch of the phase diagram was presented previously \cite{HaradaLT}
and is reproduced in Fig. \ref{PhaseDiagram}.
For the one-dimensional system,
there are several special points at which 
exact results are available \cite{Southerland075,Takhtajan,Babujian,German050},
and the complete phase diagram has been established,
except for the possibility of an intermediate phase
between the ferromagnetic phase and the dimerized phase
 \cite{Chubukov,FathSolyom,KawashimaPTP},
whose presence or absence does not have any direct consequence to
the present letter though
 \cite{Harada}.
In two dimensions
the mean-field theory \cite{ChenL1973,Papanicolaou1986} predicted 
the zero-temperature phase diagram 
as shown in the top part of Fig. \ref{PhaseDiagram}, which was confirmed by
Quantum Monte Carlo simulations \cite{HaradaK2001,HaradaK2002}.
The main focus of the present Letter is the phase
transition that occurs as we go from one dimension to two dimensions
by increasing the coupling between chains. We introduce the anisotropy 
parameter $\lambda$ such that
$J_{ij} = J$ when the bond $(ij)$ is along the chain ($x$)  to one
direction
and $J_{ij} = \lambda J$ if the bond $(ij)$ is perpendicular to the chains
(the $y$ direction). 


Our Monte Carlo simulations are based on the loop algorithm 
with non-binary loops \cite{Review}.
The system sizes we explored range from $L_x=4$ to $L_x=192$
with a fixed aspect ratio $L_x/L_y=4$.
For each data point, we typically perform a simulation 
of more than $800,000$ Monte Carlo sweeps.
The inverse temperature ranges from $\beta J = L_x/2$ to $2 L_x$
for the system size $L_x \le 128$. 
We have checked that the estimated values of quantities
discussed here do not depend on the temperature once the inverse 
temperature exceeds $L_x/J$. 
Therefore, we consider that we can regard the estimates in the present
letter as zero-temperature values.
We have measured two observables:
the quadrupolar moment and the dimerization parameter.
While the former characterizes the breaking of the spin rotational 
symmetry, the latter characterizes the breaking of the 
translational symmetry.
We define the quadrupolar moment as $Q \equiv (S^x)^2 - (S^y)^2$.
As dimerization order parameter,
we use
$
 D(\vect{r}) \equiv
 (-1)^{r_x} ( H_d(\vect{r}+\vect{e}_x) - H_d(\vect{r}) ),
$
where
$
 H_d(\vect{r}) \equiv
 {\rm diag} \left( H_{\vect{r} , \vect{r}+\vect{e}_x} \right),
$
and  $\vect{e}_x$ is the lattice unit vector in the $x$-direction.



We first look at the special point $\theta = -\pi/2$, which has 
an enlarged SU(3) symmetry.
In this case, the quadrupolar moment can be transformed to
the staggered magnetization by a `staggered' SU(3) rotation.
Because of the symmetry change exactly at $\theta = -\pi/2$ and
our previous numerical evidence in two dimensions,
we conjecture that the phase boundary between the quadrupolar 
region and the N\'{e}el region is vertical and straight exactly 
at $\theta = -\pi/2$ as depicted in Fig. \ref{PhaseDiagram}
and ends at the critical point (marked as `X' in Fig. \ref{PhaseDiagram}).
Along this line, both the quadrupolar moment and 
the staggered magnetization are finite.
In order to locate the transition point, one usually
computes the structure factors or, more conveniently, 
the Binder parameter (the dimensionless ratio of the moments of 
the order-parameter distribution function).
However, they are integrated quantities and, in the present case, 
are dominated by short-range behavior of the correlation function.
Therefore, we have used as our probe the correlation function 
between two distant points.
In particular, the ratio of the two-point correlation function 
of the largest distance ($L_x/2$) to that of a half that distance ($L_x/4$)
\cite{Tomita} is useful.
For the quadrupolar moment, we compute
$
 R_Q(L_x) \equiv 
 \left. G_Q\left(\frac{L_x}{2}\right)\ \right/ 
 \ G_Q\left(\frac{L_x}{4}\right)
$
where 
$
 G_Q(l) \equiv \left\langle Q(l\vect{e}_x) Q(\vect{0}) \right\rangle.
$
The correlation ratios for the dimerization 
$D$ are defined similarly.
%
\deffig{CorrelationRatios_050}{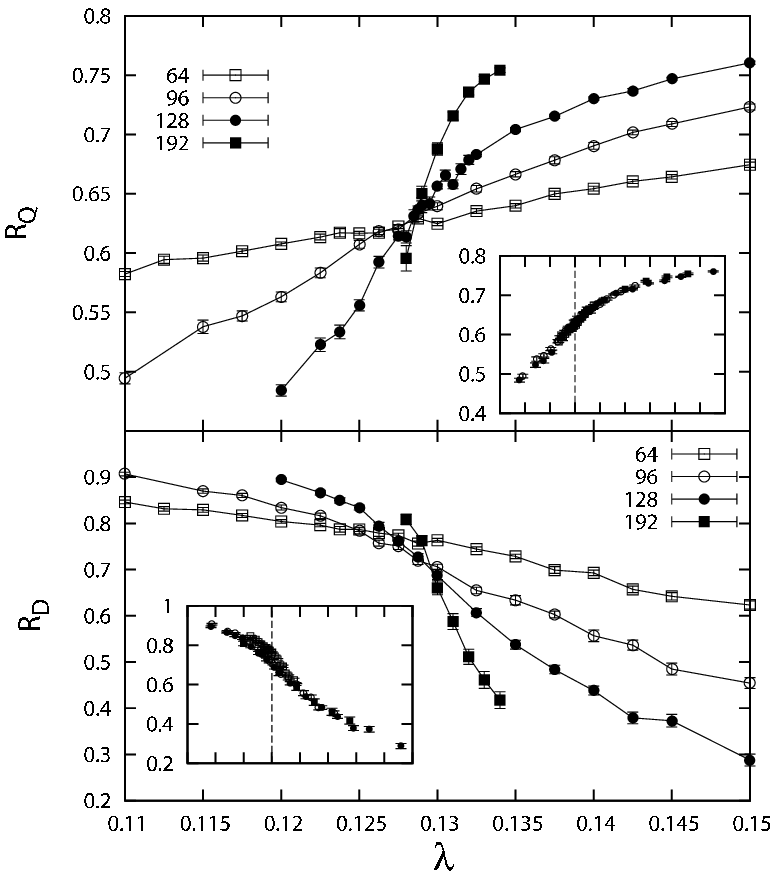}{1.00}
{ 
 The ratio of the two-point correlation function of 
 the largest distance ($=L_x/2$) to that of a half of it ($=L_x/4$)
 for the quadrupolar moment (the upper panel) and the dimerization 
 (the lower panel) at $\theta = -\pi/2$.
 The insets are finite-size-scaling plots with 
 $\lambda_c = 0.1286$,\ $y_{\rm Q} = 2.9$ (the upper panel) and
 $\lambda_c = 0.1296$,\ $y_{\rm D} = 2.54$ (the lower panel).
 The vertical dashed lines in the insets indicate the position of zero
 (the transition point).
}
%
The correlation ratio for the quadrupolar moment is shown 
in the upper panel of  Fig. \ref{CorrelationRatios_050}.
As expected, the curves for different system sizes share a common crossing
point, and it is near $\lambda \sim 0.129$.
A similar analysis is applied to the dimerization correlation function.
(See the lower panel of Fig. \ref{CorrelationRatios_050}.)
Again all the curves cross each other in the vicinity of a single point.
The location of this common crossing point is 
$\lambda \sim 0.129$ and $R_{D} \sim 0.75$.
For a more systematic analysis, we also apply finite-size-scaling 
as shown in the insets of Fig. \ref{CorrelationRatios_050}.
From the quadrupolar moment, we obtain
$
 \lambda_c = 0.1287(8),
$
\ 
$
 y_{\rm Q} = 1/\nu_{\rm Q} = 2.9(2).
$
The critical exponent, $\eta_{\rm Q}$, can be estimated from the size 
dependence of the correlation function $G_Q(L/2)$ by assuming
$G_Q(L/2) \propto L^{-(D-2+\eta_{\rm Q})}$ ($D \equiv d + z$)
at the critical point estimated above.
The results is 
$z + \eta_{\rm Q} = 1.2(1)$.
It is tempting to compare this result with the prediction of the
deconfinement critical-phenomena (DCP).
\cite{SenthilVBSF2004,SenthilBSVF2004,SenthilBSVF2005},
Although it is not totally clear whether such a comparison is 
appropriate in spite of the difference in the symmetry of the systems 
(the four-fold rotational symmetry in 
\cite{SenthilVBSF2004} and \cite{SenthilBSVF2004} and the two-fold
symmetry in the present case),
the present result for $\eta_{\rm Q}$ is certainly larger than the
value of the universality class of the three-dimensional Heisenberg model,
which is consistent with the prediction of the DCP.
A similar analysis for the dimerization yields
$
 \lambda_c = 0.1288(8),
$
\ 
$
 y_{\rm D} = 2.5(2).
$
As for $\eta_{\rm D}$,
we do not quote an estimate here because 
a power law behavior has not been observed for a large
enough range of the system size to establish a reliable estimate.

It should be kept in mind here, however, that these
analyses are based on the assumption of
a conventional critical point, which may 
not be justified in the present case.
In fact, there are reasons to expect poor quality 
of the finite-size-scaling plot,
if the present critical phenomenon is similar to the one
discussed in Refs. \cite{SenthilVBSF2004,SenthilBSVF2004,SenthilBSVF2005}.
Firstly, there must be two length scales 
(such as $\xi$ and $\xi_{\rm VBS}$ in \cite{SenthilBSVF2005}) 
diverging with different exponents.
Therefore a good scaling plot is possible only 
for system sizes greater than the shorter (but still divergent) scale.
Secondly, due to the existence of the secondary fixed point
(analogous to the U(1) spin liquid fixed point in \cite{SenthilBSVF2005})
for the renormalization flow to the VBS fixed point,
the apparent value of the scaling exponent $y$
in one phase can differ from that in the other,
which could be fatal for an ordinary finite-size-scaling
analysis.


We next look at the case $\theta = -0.55\pi$ and examine
the nature of the transition at the point `Y' in 
Fig. \ref{PhaseDiagram}.
In the previous case of $\theta = -\pi/2$, 
if we regard the ground state as the N\'eel state
rather than the spin nematic state, 
the symmetry group of the magnetic phase {\it is}
a sub-group of that of the dimerized state.
Therefore, the transition is not `unconventional' even if it
is of the second order.
This is in contrast to the case studied previously
 \cite{SenthilBSVF2004,SenthilBSVF2005} where the lattice 
rotation (by 90 degrees) is spontaneously broken in dimerized 
phase whereas it is not broken in the N\'eel state.
At $\theta = -0.55\pi$, the SU(3) invariance 
does not exist and the model has only the natural SU(2) invariance.
In the spin nematic phase, the symmetry related to the
spin space is reduced to $Z_2$ due to the spontaneous
quadrupolar moment whereas the spatial symmetry is 
the group of all the lattice translations.
In the dimerized phase, on the other hand,
the full SU(2) symmetry exists in the spin space whereas
the spatial symmetry is reduced to the sub-group
that does not contain odd translations in the $x$-direction.
Therefore, in the conventional picture of
second-order transitions, the transition between 
two such phases must be of the first order.

\deffig{CorrelationRatios_055}{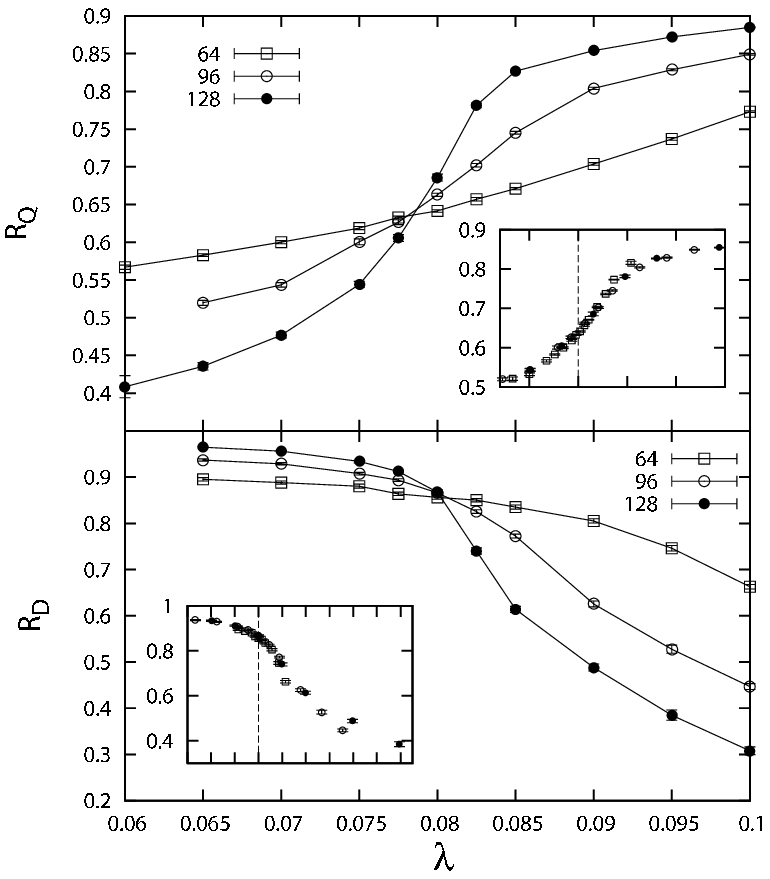}{1.00}
{ 
 Similar to Fig. \ref{CorrelationRatios_050} but for $\theta = -0.55 \pi$.
 The parameters used for the insets are
 $\lambda_c = 0.0788$,\ $y_{\rm Q} = 2.9$ (the upper panel) and
 $\lambda_c = 0.0800$,\ $y_{\rm D} = 2.8$ (the lower panel).
}

Recently, however, the possibility of second-order transitions 
was discussed extensively and a new paradigm, 
namely the deconfinement critical phenomenon, was proposed.
\cite{SenthilVBSF2004,SenthilBSVF2004,SenthilBSVF2005},
In this scenario, the VBS state is viewed as a state in which
$S=1/2$ quasi-particles are confined by the potential
that is sustained by the spontaneous VBS order
to form $S=1$ elementary excitations.
As the quantum phase transition is approached, the VBS order
is weakened and so is the confinement potential.
Eventually, the quasi-particles are deconfined at the critical point,
giving rise to quasi-long-range order.
It was also predicted that the two phases with apparently 
unrelated symmetries may have a direct second-order 
transition due to this mechanism.
For this hypothetical class of phase transitions,
there are efforts to find an example 
that exhibits such a transition \cite{Sandvik2006}.
However, so far no phase transition of any model is confirmed
beyond reasonable doubt to match the theoretical description of DCP.
In particular, it was shown in Ref. \cite{Kuklov2006} that
the DCP may not generically exist in models with U(1) symmetry.

In Fig. \ref{CorrelationRatios_055} we plot the correlation ratio 
at $\theta = -0.55\pi$, 
We can see that both sets of curves cross near $\lambda = 0.078$,
an indication of a single phase transition between two phases.
For the quadrupolar moment, the finite-size-scaling analysis yields
$ 
 \lambda_c = 0.079(1),\ y_{\rm Q} = 2.8(2).
$
From the size dependence of the correlation, we obtain
$z + \eta_{\rm Q} = 1.25(20)$.
For the dimerization, on the other hand, we obtain
$ 
 \lambda_c = 0.080(1),\ y_{\rm D} = 2.8(2).
$
The system size dependence of the correlation ratio
is not in perfect accordance 
with a first-order transition:
if the transition is of the first order,
the correlation ratios for both the quantities must be 1
simultaneously at the transition point.
However, the ratio at the transition is almost constant
and shows no visible trend towards unity.
The correlation function decreases
algebraically at the transition at least up to a 
distance of $128$ lattice spacings.

Finally, in Fig. \ref{E_vs_Lambda_055}, 
we plot the energy per site
as a function of $\lambda$.
The system size dependence of the energy per site 
is small and no significant
structure visible in this figure, in particular near
the transition point $\lambda \sim 0.079$.
It is still possible that the present transition is a
very weak first-order transition of which the
jump in the first derivative is impossible
to observe within the present computational accuracy.
Another open possibility is that the present phase transition 
is actually two second-order transitions very close to each other.
While we cannot completely rule out these possibilities,
we have obtained no evidence that positively supports 
either possibility.

\deffig{E_vs_Lambda_055}{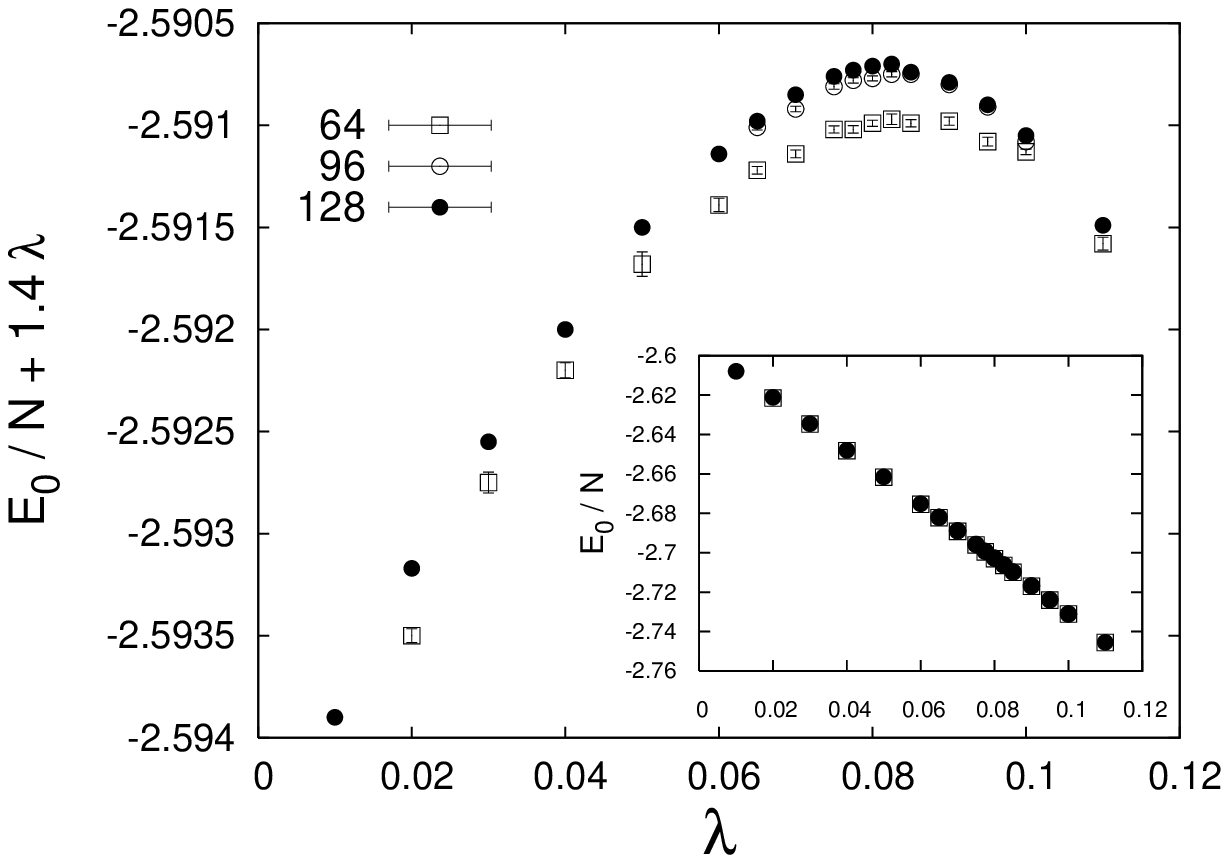}{0.65}
{
 The ground-state energy per site for $\theta = -0.55\pi$,
 after subtraction of the dominant 
 linear contribution  $1.4 \lambda$, highlighting the absence of 
 any cusp in the energy near the transition point
 ($\lambda \sim 0.079$). 
 The inset shows the raw data before subtraction.
}

The last and the most interesting scenario consistent with the present
numerical evidences is a single second-order transition.
If this is the case, the transition is unconventional as discussed above,
and may correspond to the predicted DCP scenario which is absent
in generic U(1) symmetric models.\cite{Kuklov2006}
Because of the difference in the spatial 
symmetry of the model, namely, the lack of the lattice rotational 
symmetry in the present model,
it is not clear how the phase transition studied in this Letter 
is related to the specific mechanism of deconfined criticality
that involves monopole-type excitations.
However, the generic scenario proposed in the previous studies 
\cite{SenthilVBSF2004,SenthilBSVF2004,SenthilBSVF2005}, 
in which a secondary fixed point exists and 
the translational-invariance-breaking field becomes relevant 
at this point, may still apply to the present case.
While the computations shown in this Letter are already fairly extensive 
(roughly 20 cpu$\cdot$years), even larger-scale computations 
are being prepared for in order to single out the correct scenario 
for the transition in the SU(2)-symmetric model and 
and to clarify its mechanisms.


We are grateful to F.~Alet,  A.~L\"{a}uchli, S.~Trebst and C.~Wu
for valuable and stimulating discussions.
KH's work was supported by the 21st Century COE Program
of Monkasho, Japan.
MT acknowledges the Aspen Center for Physics.
Preliminary studies were carried out on Beowulf
clusters Asgard and Hreider at ETH Z\"urich. 
The main part of the computation was performed on SGI Altix3700/1280 
at Supercomputer Center, Institute of Solid State Physics, 
University of Tokyo.


\begin{thebibliography}{99}

\bibitem{ColdAtoms}
S.~Inouye, {\it et al.}, Nature {\bf 392}, 151 (1998).

\bibitem{OpticalLattice}
M.~Greiner, {\it et al.}, Nature {\bf 415}, 39 (2002).

\bibitem{BatrouniETAL2002}
G.~G.~Batrouni, {\it et al.}, Phys.\ Rev.\ Lett.\ {\bf 89}, 117203 (2002).

\bibitem{Kashurnikov2002}
V.A. Kashurnikov, N.V. Prokof'ev, and B V. Svistunov, Phys. Rev. A {\bf 66}, 
031601(R) (2002).

\bibitem{Wessel2004}
S.~Wessel, F.~Alet, M.~Troyer and G.~G.~Batrouni, 
Phys. Rev. A {\bf 70}, 053615 (2004).

\bibitem{Bergkvist2004}
S. Bergkvist, P. Henelius, and A. Rosengren, Phys. Rev. A {\bf 70},
053601 (2004).

\bibitem{Yip} 
  S. K. Yip,
  Phys. Rev. Lett. {\bf 90}, 250402 (2003).

\bibitem{Imambekov}  
  A.~Imambekov, M.~Lukin and E.~Demler,
  Phys.\ Rev.\ A {\bf 68}, 063602 (2003).

\bibitem{SenthilVBSF2004}  T.~Senthil, A.~Vishwanath, L.~Balents, S.~Sachdev, and
                           M.~P.~A.~Fisher,
                           Science {\bf 303}, 1490 (2004).

\bibitem{SenthilBSVF2004}  
T.~Senthil, L.~Balents, S.~Sachdev, A.~Vishwanath, and
                           M.~P.~A.~Fisher,
                           Phys.\ Rev.\ B {\bf 70}, 144407 (2004).

\bibitem{SenthilBSVF2005}
T.~Senthil, L.~Balents, S.~Sachdev, A.~Vishwanath, and M.~P.~A.~Fisher,
                           J.\ Phys.\ Soc.\ Jpn.\ {\bf 74} Suppl., 1 (2005).

\bibitem{HaradaLT} 
  K.\ Harada, N.\ Kawashima and M.\ Troyer,
  to appear in the proceedings of {\it The 24th International Conference
  on Low Temperature Physics (LT24)} (The AIP Conference Proceedings Series).

\bibitem{Southerland075} 
  B. Sutherland,
  Phys. Rev. B {\bf 12}, 3795 (1975).

\bibitem{Takhtajan} 
  L. A. Takhtajan, Phys. Lett. {\bf 87} A, 479 (1982).

\bibitem{Babujian}
  H. M. Babujian, Phys. Lett. {\bf 90} A, 479 (1982).

\bibitem{German050} 
  A. Kl\"umper,
  Euro. Phys. Lett. {\bf 9}, 815 (1989).

\bibitem{KawashimaPTP} 
  N. Kawashima,
  Prog.\ Theor.\ Phys.\ (Suppl.) {\bf 145}, 138 (2002).

\bibitem{Chubukov} 
  A. V. Chubukov,
  Phys.\ Rev.\ B {\bf 43}, 3337 (1991).

\bibitem{FathSolyom} 
  G. F\'ath and J. S\'olyom,
  Phys.\ Rev.\ B {\bf 51}, 3620 (1995).

\bibitem{Harada}
  In our simulations,
  we observe an apparent transition point (between $\theta = -0.75\pi$ and
  $\theta = -0.5\pi$) at which the spontaneous dimerization 
  seems to vanish. However, this apparent transition point is systematically 
  closer to $\theta = -0.75\pi$ for larger systems,
  consistent with the absence of the intermediate phase and
  an exponentially-diverging correlation length at $\theta = -0.75\pi$.

\bibitem{ChenL1973}        H.~H.~Chen and P.~M.~Levy,
                           Phys.\ Rev.\ B {\bf 7}, 4267 (1973).

\bibitem{Papanicolaou1986} N.~Papanicolaou,
                           Phys.\ Lett.\ A {\bf 116}, 89 (1986);
                           Nucl.\ Phys.\ {\bf B305}[FS23], 367 (1988).

\bibitem{HaradaK2001}      
  K.~Harada and N.~Kawashima,
  J.~Phys.~Soc.~Jpn.\ {\bf 70}, 13 (2001).

\bibitem{HaradaK2002}      
  K.~Harada and N.~Kawashima,
  Phys.\ Rev.\ B {\bf 65}, 52403 (2002).

\bibitem{Review}           N.~Kawashima and K.~Harada,
                           J.\ Phys.\ Soc.\ Jpn.\ {\bf 73}, 1379 (2004).

\bibitem{Tomita} 
Y. Tomita and Y. Okabe,
Phys.\ Rev.\ B {\bf 66}, 180401(R) (2002).


\bibitem{Sandvik2006}
A.~W.~Sandvik and R.~G.~Melko,
arXiv,cond-mat/0604451.

\bibitem{Kuklov2006}
A.~B.~Kuklov, N.~V.~Prokof'ev, B.~V.~Svistunov and M.~Troyer,
arXiv,cond-mat/0602466.

\end{thebibliography}
\end{document}